\documentclass[epj,twocolumn]{webofc}
\usepackage[varg]{txfonts}   % Web of Conferences font
\usepackage{graphicx}

\wocname{epj}
\woctitle{Seismology of the Sun and the Distant Stars 2016}
\begin{document}
%***********************************************************************
\title{Global helioseismology (WP4.1): From the Sun to the stars \& solar analogs}
%\subtitle{from the Sun to the stars \& solar analogs}

\author{\firstname{Rafael A.} \lastname{Garc\'\i a}\inst{1}\fnsep\thanks{\email{rafael.garcia@cea.fr}}
        % etc.
}

\institute{Laboratoire AIM Paris-Saclay, CEA/DRF-CNRS-Univ. Paris Diderot; IRFU/SAp, Centre de Saclay, 91191 Gif-sur-Yvette, France 
}

%-----------------------------------------------------------------------
\abstract{%idl
Sun-as-a star observations put our star as a reference for stellar observations. Here, I review the activities in which the SPACEINN global seismology team (Working Package WP4.1) has worked during the past 3 years. In particular, we will explain the new deliverables available on the SPACEINN seismic+ portal. Moreover, special attention will be given to surface dynamics (rotation and magnetic fields). After characterizing the rotation and the magnetic properties of  around 300 solar-like stars and defining proper metrics for that, we use their seismic properties to characterize 18 solar analogues for which we study their surface magnetic and seismic properties. This allows us to put the Sun into context compared to its siblings.
}
\maketitle
%
%-----------------------------------------------------------------------
\section{Introduction}
\label{intro}
The objective of the global seismology team inside the SPACEINN collaboration, namely Working Package 4.1, is to discuss the current problems in global seismology of the Sun and solar-analogue pulsating stars. Naturally, attending to the type of observations, i.e. Sun-as-a-star or imaged ones, the work and the discussions related were divided in low and high-degree modes. In this proceedings I will concentrate on the activities of the group concerning the low-degree modes obtained in Sun-as-a star observations and in the studies of solar analogues. 

During the past three years, the discussion of high-degree global modes were focused on the use of different fitting codes and the differences obtained when several datasets of different instruments were used. Some of the works done in this direction are summarized in \cite[e.g.][]{2014AAS...22421813K,2016SPD....4710701K}. Briefly, it has been found that some of the p-mode parameters (for example the mode asymmetries) are inconsistent when different fitting methods are used, while they are consistent when a single methodology is used in different datasets, e.g., with GONG \cite{HarHil1996} or with MDI and HMI \cite{1995SoPh..162..129S,2012SoPh..275..207S}. Moreover, inversions of the internal rotation profile showed that the solar magnetic cycle 24 is different from cycle 23. The work with different datasets and methods evidenced that the twist appearing at high latitudes with time in several helioseismic variables is very likely to be a numerical artifact \cite[for more details see][]{2016SPD....4710701K}. 

Concerning the low-degree modes, the two main axes of the work in the WP4.1 were concentrated on the development of new statistical tools to study these modes and the study of the temporal evolution of their properties  during the evolution of the solar magnetic activity cycle. The solar results were placed into a stellar context by comparing them to observations of other solar analogue stars observed by the NASA \emph{Kepler} mission \cite{2010Sci...327..977B}.

%-----------------------------------------------------------------------
\section{Real and simulated solar time series}
\label{sec-Calib}

A huge effort has been led by the WP4.1 in order to provide properly calibrated datasets obtained from the Sun-as-a-star instruments on board the Solar and Heliospheric Observatory (SoHO \cite{DomFle1995}), as well as from the Mark-I instrument, a solar spectrophotometer located and operated at Observatorio del Teide (Tenerife, Canary Islands, Spain). It provides precise radial velocity observations of the Sun-as-a-star\footnote{http://www.spaceinn.eu/data-access/mark-i-data-archive/} at the Potassium KI 7699 \AA~absorption solar line. Observations extend from 1976 to 2012 although only the summer campaigns of 1976 to 1983 are available.

Photometric light curves of the Sun Photometers (SPM) of the Variability of solar IRradiance and Gravity Oscillations instrument (VIRGO \cite{1995SoPh..162..101F}) are now available at the SPACEINN portal\footnote{http://www.spaceinn.eu/data-access/calibrated-sohovirgospm-data/}. These are 60~s cadenced time series starting on April 11, 1996 of the three channels of 5~nm bandwidth centered at 402~nm (blue), 500~nm (green) and 862~nm (red). These light curves are corrected from outliers and then filtered with a high-pass filter running mean to remove unwanted low-frequency trends. An additional correction is also applied whenever possible in order to correct for the so-called SPM ``attractor'' as explained in \cite{2002SoPh..209..247J}. 
 
Doppler velocity time series observed by the Global Oscillations at Low Frequency (GOLF \cite{GabGre1995}) instrument are also available in the SPACEINN portal\footnote{http://www.spaceinn.eu/data-access/calibrated-soho-golf-frequencies/}. Observations obtained from the blue or the red wing of the Sodium doublet \cite{2004ESASP.559..432G}, each one with a different sensitivity to the solar disk \cite{1998A&AS..128..389G}, have been calibrated following the methods explained in \cite{2005A&A...442..385G} and the two independent channels have been averaged into one single time series.  The starting date is at midnight of April 11, 1996 and the sampling rate is 60~s in order to follow the same cadence than the other two helioseismic instruments on board SoHO. 

Numerical simulations have also been created and are available on demand. All the information can be found in the SPACEINN portal\footnote{http://www.spaceinn.eu/data-access/wp4/simul/}. Three different type of simulations were created:

\begin{enumerate}
\item Spherical-shell simulations of magnetoconvection.
\item Numerical Simulations of linear waves/modes in complex media.
\item Box simulations of compressible magnetoconvection.
\end{enumerate}

%.......................................................................
\section{Fitting low-degree solar modes}
\label{sec-Fit}
\begin{figure}[!htb]
\centering
\includegraphics[width=\hsize,clip]{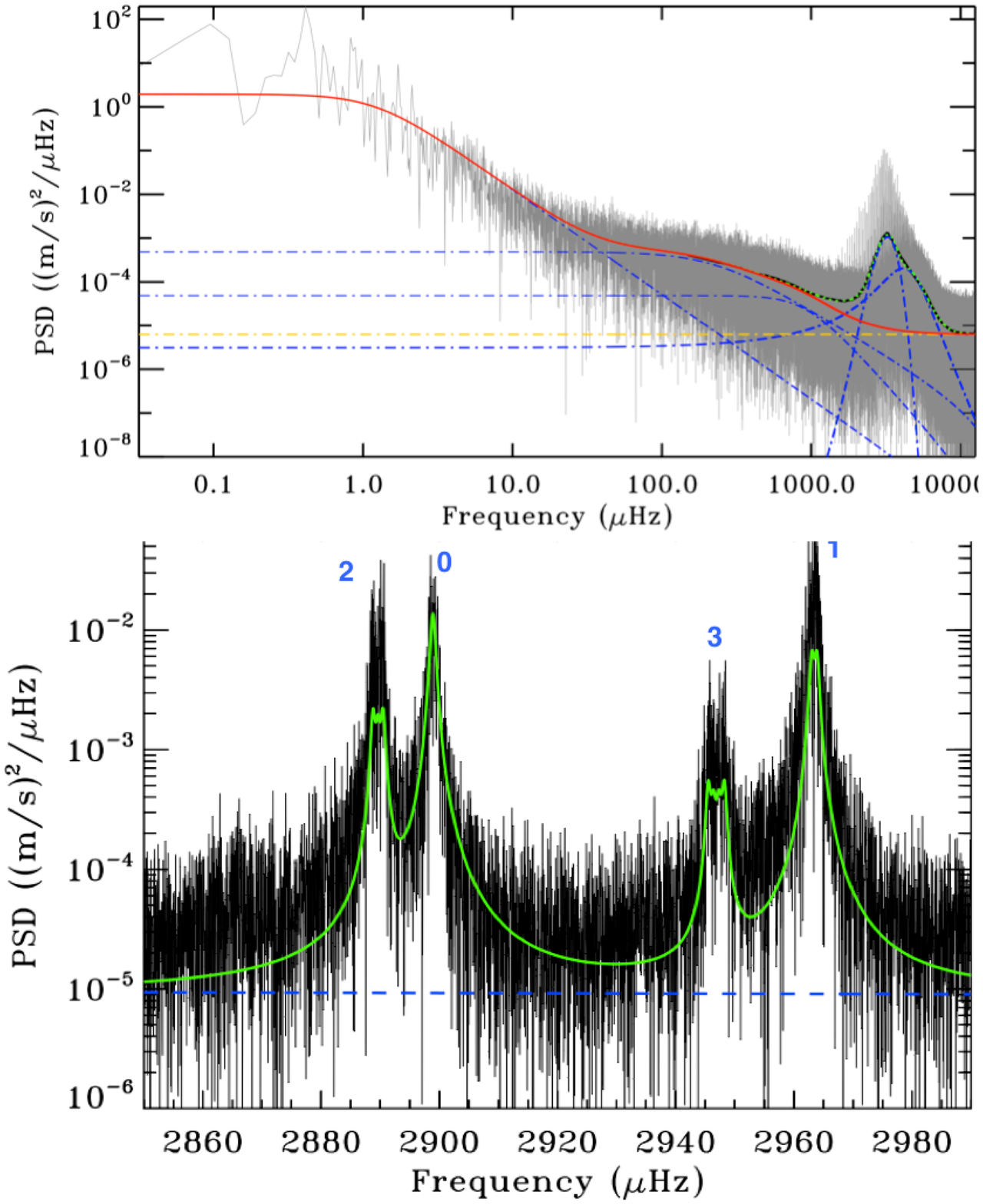}
\caption{Application of the DIAMONDS asteroseismic Bayesian global-fitting tool \cite{2014A&A...571A..71C} to the solar case. On the top panel, result of the fitting of the background and the power excess of GOLF velocity time series using a model of three Harvey-like profiles, 2 Gaussian functions and a flat photon-noise contribution. On the bottom panel, zoom of the region 2850 to 2990 $\mu$Hz in black and the result of the fit of modes $\ell$=0 to 3 in green. The horizontal dashed line represents the background level.
%PSD of the \emph{Kepler} target KIC~3733735 observed continuously during three years with a sampling rate of 1 minute. The physical phenomena responsable of the features in each region of the PSD are indicated: photon noise, acoustic oscillations, convection (granulation), and surface rotation induced by spot modulation of the light curve. Adapted from \cite{2015EAS....73..193G}. 
}
\label{fig-1}       % Give a unique label
\end{figure}

In helioseismology, it is common practice to characterize the acoustic modes by fitting single modes using a frequentist approach at low frequencies and pairs of even or odd modes at higher frequencies \cite[e.g.][]{AppGiz1998,1998MNRAS.300.1077C,2001SoPh..200..361G,2014MNRAS.439.2025D}. But with the development of asteroseismology and due to the natural complications to properly identify the modes, global fittings were established as standard methods \cite[e.g.][]{2008A&A...488..705A,2009A&A...506...41G,2009A&A...506...51B,2010A&A...518A..53M,2011ApJ...733...95M,2011A&A...534A...6C}. Moreover, bayesian fitting techniques were soon adopted in asteroseismology \cite[e.g.][]{2008AN....329..485A,Benomar2009a,2010CoAst.161...39W,2011A&A...527A..56H,2014A&A...571A..71C,2016MNRAS.456.2183D} and also tested on the Sun \cite{2010MNRAS.406..767B}. Although the difference between the bayesian and the frequentist approach in the case of the Sun could be neglected \cite[e.g.][]{2015MNRAS.454.4120H}, we decided to adapt the high-DImensional And multi-MOdal NesteD Sampling asteroseismic Bayesian global-fitting tool \cite[DIAMONDS,][]{2014A&A...571A..71C} to the solar case\footnote{http://www.spaceinn.eu/data-access/wp4/tools-to-extract-low-frequency-mode-frequencies-d4-4/}. An example of the application of the code to the GOLF velocity time series can be seen in figure~\ref{fig-1}. Because of the particularities of the GOLF measurements, the p-mode power excess has been fitted using a double Gaussian following \cite{2008A&A...490.1143L}. A full analysis of the GOLF and VIRGO/SPM datasets using this tool will be soon available (Corsaro et al. in preparation).

%++++++++++++++
%++++++++++++++

A lot of discussions and efforts were employed to study low-frequency p modes and g modes in the Sun. Although the signature of the g-mode period spacing was found using GOLF observations \cite{2007Sci...316.1591G,2008AN....329..476G}, it has been impossible to unambiguously detect individual g modes with a high confidence level \cite{2010A&ARv..18..197A,GarciaHofA2010,2011JPhCS.271a2046G}. Special effort has been spent to find better ways to combine contemporary datasets as it was already outlined in \cite{2010MNRAS.406..767B}. Proper covariant matrices have been developed (Broomhall et al. in preparation) but the results have not allowed us to find solar g modes so far. 

%.......................................................................
\section{Photospheric magnetic activity proxy}

\begin{figure}[!htb]
\centering
\includegraphics[width=\hsize,clip]{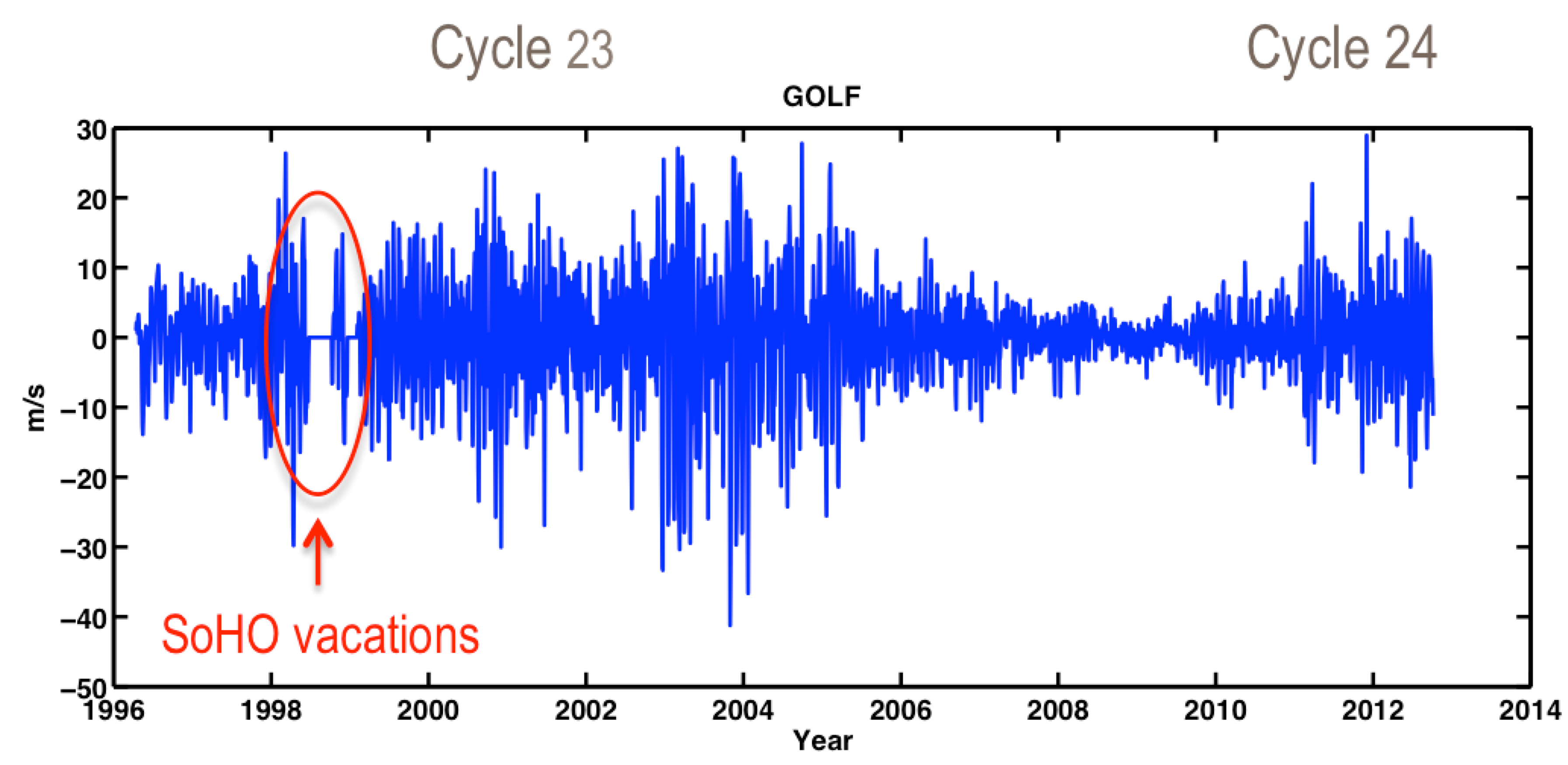}
\caption{Surface magnetic activity proxy computed form the GOLF observations, $S_{\rm{vel}}$, as a function of time from the raising phase of solar magnetic cycle 23 to the maximum of cycle 24. The so-called SoHO vacation periods of 1998 and early 1999 are indicated. 
}
\label{fig-2}       % Give a unique label
\end{figure}

The quasi continuous space observations required to perform helio- and astero-seismic studies can be used to track the evolution of the photospheric magnetic field \cite[e.g.][]{2009ASPC..416..529M,2009A&A...506...33M,2011ApJ...732L...5C,2013A&A...549A..12M,2013A&A...550A..32M} as well as in the sub-photospheric layers through seismology \cite{2010Sci...329.1032G,2011A&A...530A.127S,2016A&A...589A.118S,2016A&A...589A.103R,2016arXiv161102029K}. In the SPACEINN framework, we have extensively studied the photospheric magnetic activity proxies obtained from photometric observations, $S_{\rm{ph}}$ \cite[e.g.][]{2014A&A...562A.124M,2014A&A...572A..34G} and Doppler velocity time series $S_{\rm{vel}}$. Briefly they track the standard deviation of subseries of a length proportional to the rotation period in order to ensure that the periodicities are related to stellar spots and hence, to magnetic activity.  An example of the solar $S_{\rm{vel}}$ measured using the GOLF data is given in figure~\ref{fig-2}. 

%++++++++++++++
%++++++++++++++

\begin{figure}[!htb]
\centering
\includegraphics[width=\hsize,clip]{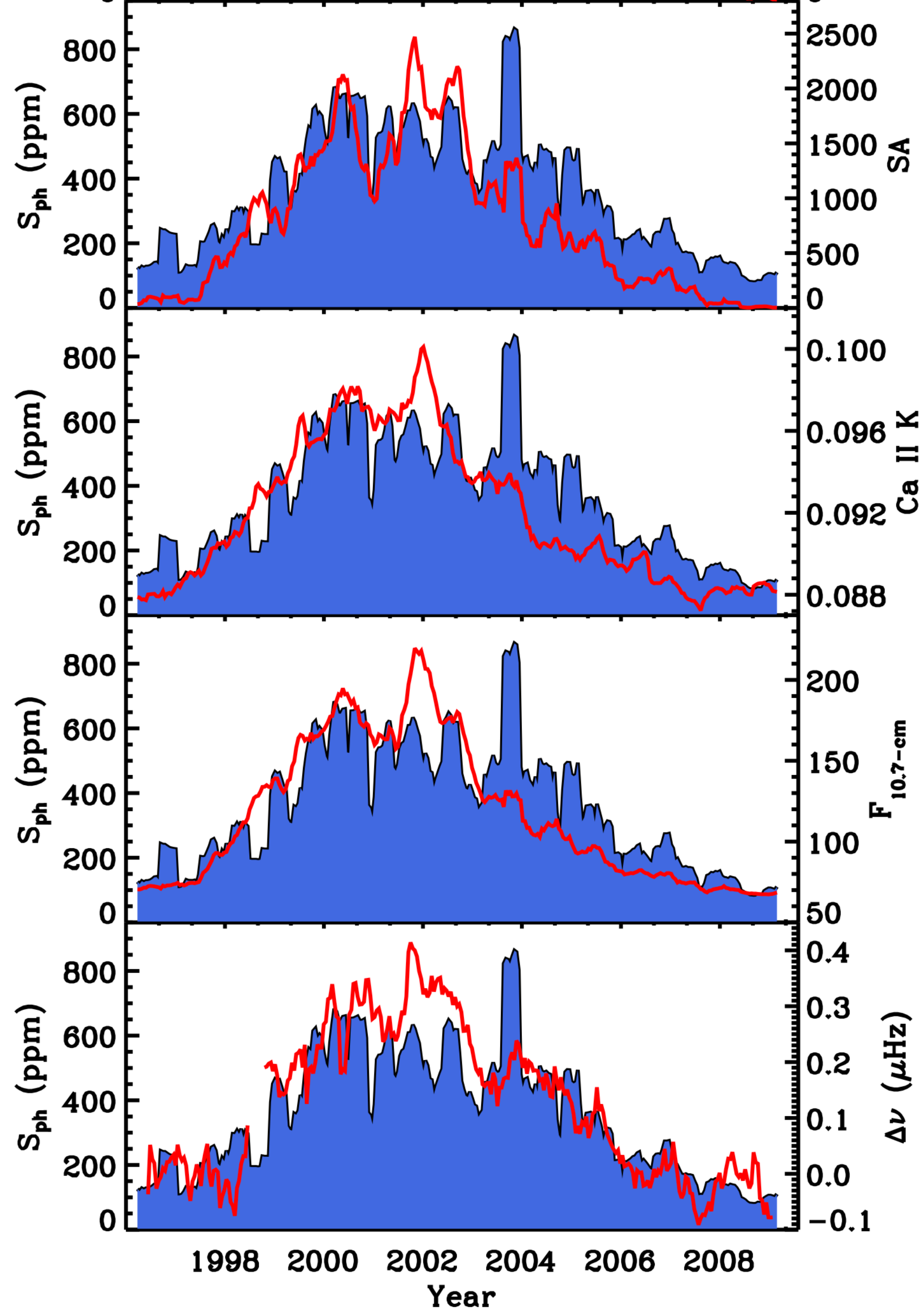}
\caption{Evolution with time of the photospheric magnetic activity proxy $S_{\rm{ph}}$ measured from VIRGO/SPM (blue filled curve) compared to some standard solar magnetic activity proxies (red solid lines). From top to bottom: the right-hand y axis corresponds to the sunspot area (SA); the Ca II K-line emission index in \AA; the 10.7-cm radio flux ($F_{10.7-\rm{cm}}$) in $10^{-22}$ s$^{-1}$ m$^{-2}$ Hz$^{-1}$; and the frequency shifts $\Delta\nu$ of the acoustic low-degree p modes in $\mu$Hz. Adapted from \cite{2016arXiv161000990S}.
}
\label{fig-3}       % Give a unique label
\end{figure}

In particular, we have used the Sun as a reference (Salabert et al. in preparation) and we have compared $S_{\rm{ph}}$ and $S_{\rm{vel}}$ to many other solar magnetic proxies \cite{2015SoPh..290.3095B}. An example of these comparisons is shown in figure~\ref{fig-3}.
%++++++++++++++
%++++++++++++++

Four time series of the $S_{\rm{ph}}$ corresponding to the VIRGO/SPM blue, green, red channels, and the \emph{Kepler}-like composite as well as one time series of the GOLF 
$S_{\rm{vel}}$ are available at the SPACEINN portal\footnote{http://www.spaceinn.eu/data-access/photospheric-solar-activity-index-virgospm-sph/}. These files are updated in multiples of 5x27= 135 days because, as it has been demonstrated for the Sun and other stars \cite{2014JSWSC...4A..15M}, a length of five times the rotation period provide the best compromise to study the evolution of the magnetic activity cycle. These solar datasets have been computed from the raw time series of VIRGO/SPM and GOLF applying the calibration procedures developed for \emph{Kepler} \cite{2011MNRAS.414L...6G} using a high-pass filter with a cut-off period at 60 days.
%.......................................................................
\section{Comparison of solar magnetic activity cycles 23-24}

\begin{figure}[!htb]
\centering
\includegraphics[width=\hsize,clip]{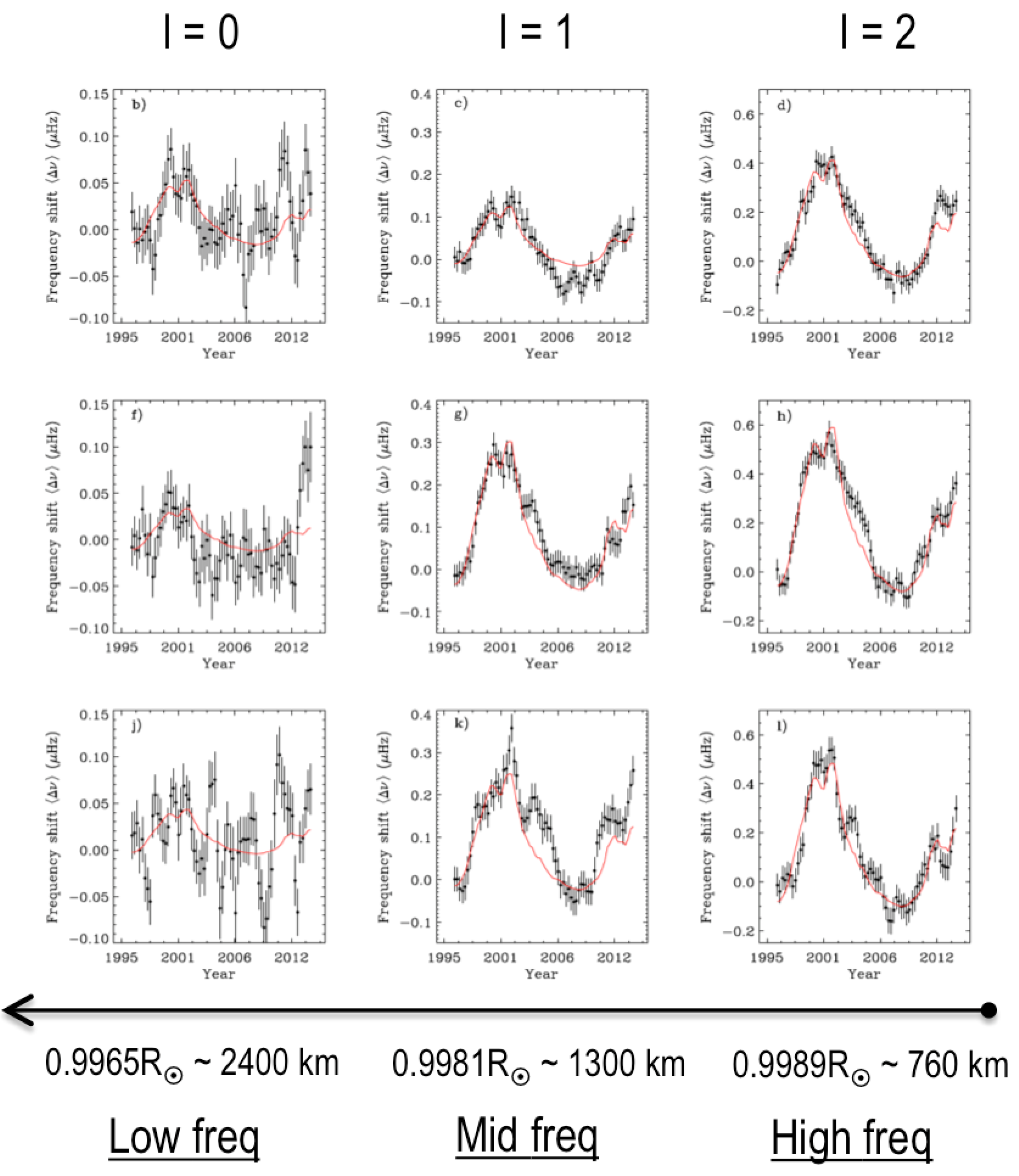}
\caption{Temporal variations of the frequency shifts in $\mu$Hz averaged over the modes $\ell$= 0, 1, and 2,  calculated for four different frequency ranges (low, middle and high frequencies), each one sensitive to a different depth. The associated 10.7-cm radio flux, F10.7, scaled to match the rising phase and the maximum of Cycle 23 is shown as a proxy of the solar surface activity (solid line). Adapted from \cite{2015A&A...578A.137S}.
}
\label{fig-4}       % Give a unique label
\end{figure}

A particular effort has been done to study the magnetic activity cycle of the Sun and to better understand the differences between cycles 23 and 24 that we know had a particularly unexpected long magnetic activity minium  \cite[e.g.][]{Livingston12,2014SSRv..186..191B}. Thanks to helioseismology, it si now possible to ``see'' inside the Sun below the photosphere and study the changes during the evolution of the solar cycle \cite[e.g.][]{2012ApJ...758...43B}. An example of the frequency shifts of modes of degrees $\ell$=0, 1, and 2 are shown in figure~\ref{fig-4}. They have been calculated at three different depths beneath the photosphere at ($\sim$2400, $\sim$1300, and $\sim$760 km) \cite{2015A&A...578A.137S}. The low-frequency modes show nearly unchanged frequency shifts between Cycles 23 and 24, with a time evolving signature of the quasi-biennial oscillation \cite{2010ApJ...718L..19F,2012A&A...539A.135S}, which is particularly visible for the quadrupole component revealing the presence of a complex magnetic structure. The modes at higher frequencies show frequency shifts 30\% smaller during Cycle 24, which is in agreement with the decrease observed in the surface activity between Cycles 23 and 24. 

%++++++++++++++
%++++++++++++++
The frequency tables associated with the work described in  \cite{2015A&A...578A.137S} are now available at the SPACEINN portal\footnote{http://www.spaceinn.eu/data-access/calibrated-soho-golf-frequencies/}. A total of 69 non-independent one-year frequency tables of modes $\ell$=0, 1, 2, and 3 are available. They were computed using 365 days time series (with a four-time overlap of 91.25 days) to avoid any perturbation induced by the 1-year orbital motion of the SoHO spacecraft and covering a total of 18 years starting in April 1996.

%.......................................................................
\section{Sellar Activity: Solar analogues}
One of the most important questions we have to answer in astrophysics is if the Sun is a typical star or not attending to its photospheric properties, in particular, to its activity. Indeed, the amount of magnetic activity in a star is crucial for the development of life. Therefore, it is important to be able to compare the Sun to its closest siblings as it has been shown that when a wider comparison is done in terms of the length of activity cycles and surface rotation rates, the Sun seems to be a particular star between the so-called active and inactive branches \cite{2007ApJ...657..486B}.

\begin{figure}[!htb]
\centering
\includegraphics[width=\hsize,clip]{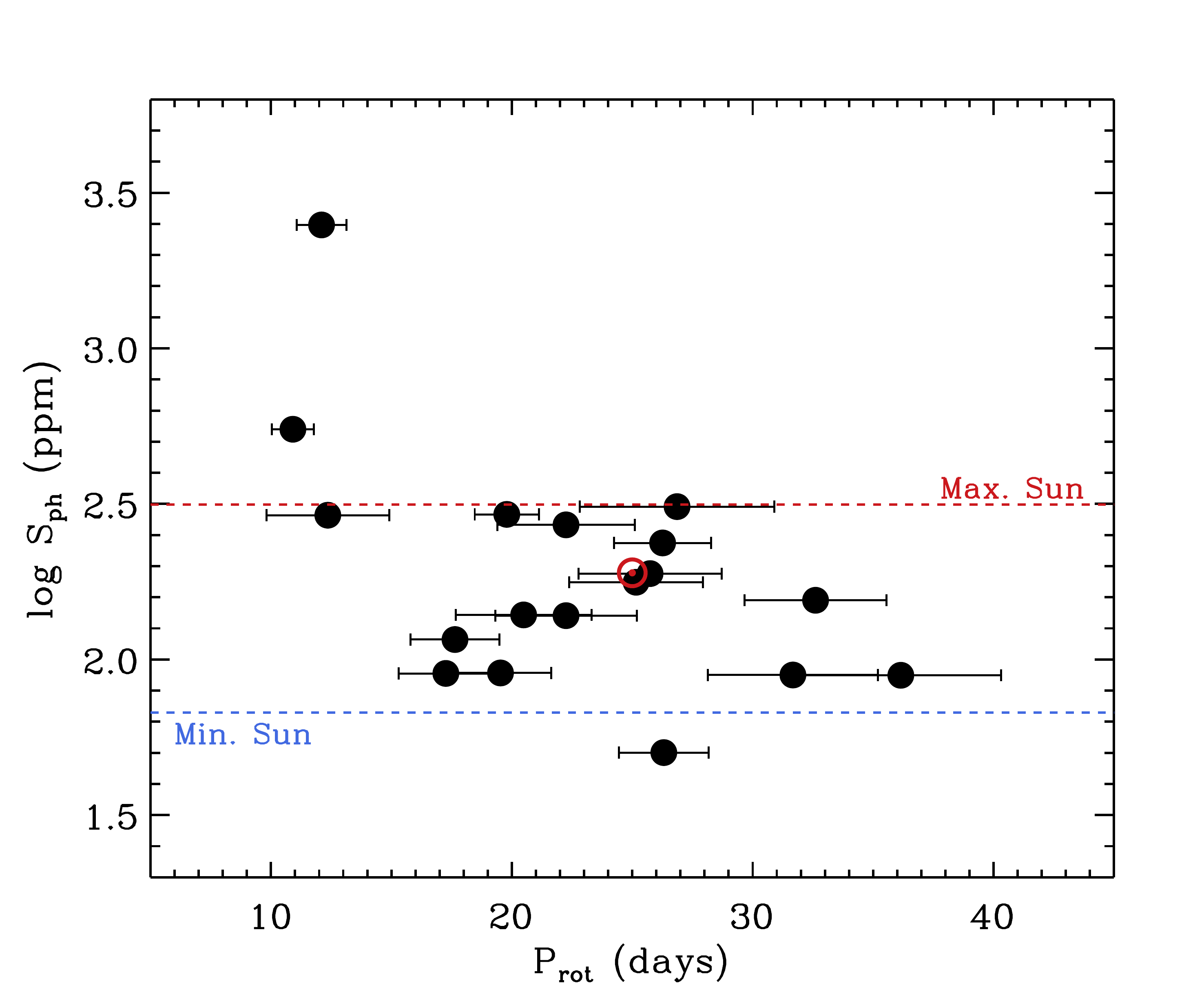}
\caption{Photospheric magnetic activity index, $S_{\rm{ph}}$ (in ppm), as a function of the rotational period, $P_{\rm{rot}}$ (in days from \cite{2014A&A...572A..34G}), of the 18 seismic solar analogs observed with the \emph{Kepler} satellite. The mean activity level of the Sun, calculated from the VIRGO/SPM observations, is represented for a rotation of 25 days with its astronomical symbol, and its mean activity levels at minimum and maximum of the 11-year cycle are represented by the horizontal dashed lines. Adapted from \cite{2016arXiv160801489S}.
}
\label{fig-5}       % Give a unique label
\end{figure}

However, we know that stars exhibiting pulsations has low surface magnetism because the latter inhibits the oscillation modes \cite{2009A&A...506...33M,2010Sci...329.1032G,2011ApJ...732L...5C,2014ApJ...785....5G}. Therefore, the first step is to compare the Sun to solar-type pulsating analogs. This has been done by several members of the WP4.1 for 18 solar analogs observed by \emph{Kepler} \cite{2016arXiv160801489S}. It has been found that the photospheric activity levels of 15 of the solar analogs are comparable to the range between the minimum and the maximum of the solar magnetic activity during the solar cycle (see figure~\ref{fig-5}). 

%++++++++++++++
%++++++++++++++

The two stars in figure~\ref{fig-5} with a higher $S_{\rm{ph}}$  correspond to the youngest stars in the sample. Hence, it is normal to have higher surface magnetic activity compared to the Sun. One star with a rotation period comparable to the Sun (i.e. with an age similar to the Sun as expected from gyro-chronology), is observed to have a photospheric activity slightly lower than the Sun at its minimum. This could be a consequence of a tilted star compared to the line of sight as the photospheric activity proxy depends on the stellar inclination angle. A similar picture can be drawn when a chromospheric proxy, the Ca K lines, are used to study the magnetism of these stars. Again, inside the error bars, all apart for the two youngest ones have activity levels inside the solar range. 

The youngest star in the sample, KIC~10644253 (1.1 $\pm$ 0.2  Gyr, \cite{2014ApJS..214...27M}), has been extensively analyzed as a possible precursor of our Sun. A magnetic activity modulation of $\sim$1.6 years has been measured in the $S_{\rm{ph}}$ as well as in the frequency shifts \cite{2016A&A...589A.118S}, and in the p-mode amplitudes \cite{2016arXiv161102029K}. This modulation could be analogous to what has been found by \cite{2015ApJ...812...12E} in the Mount Wilson star HD 30495, having very close stellar properties and falling on the inactive branch reported by \cite{2007ApJ...657..486B}. Interestingly, some discrepancies are seen when compared the seismic results with the photospheric proxy.  Because this star seems to have a low inclination angle with a weighted average value of i = 23 $\pm$ $6^o$, it is conceivable that the regions of high activity are largely confined to the nearly out-of-sight hemisphere of the star where the discrepancy between the activity proxies are the largest \cite{2016arXiv161102029K}. 

\begin{figure}[!htb]
\centering
\includegraphics[width=\hsize,clip]{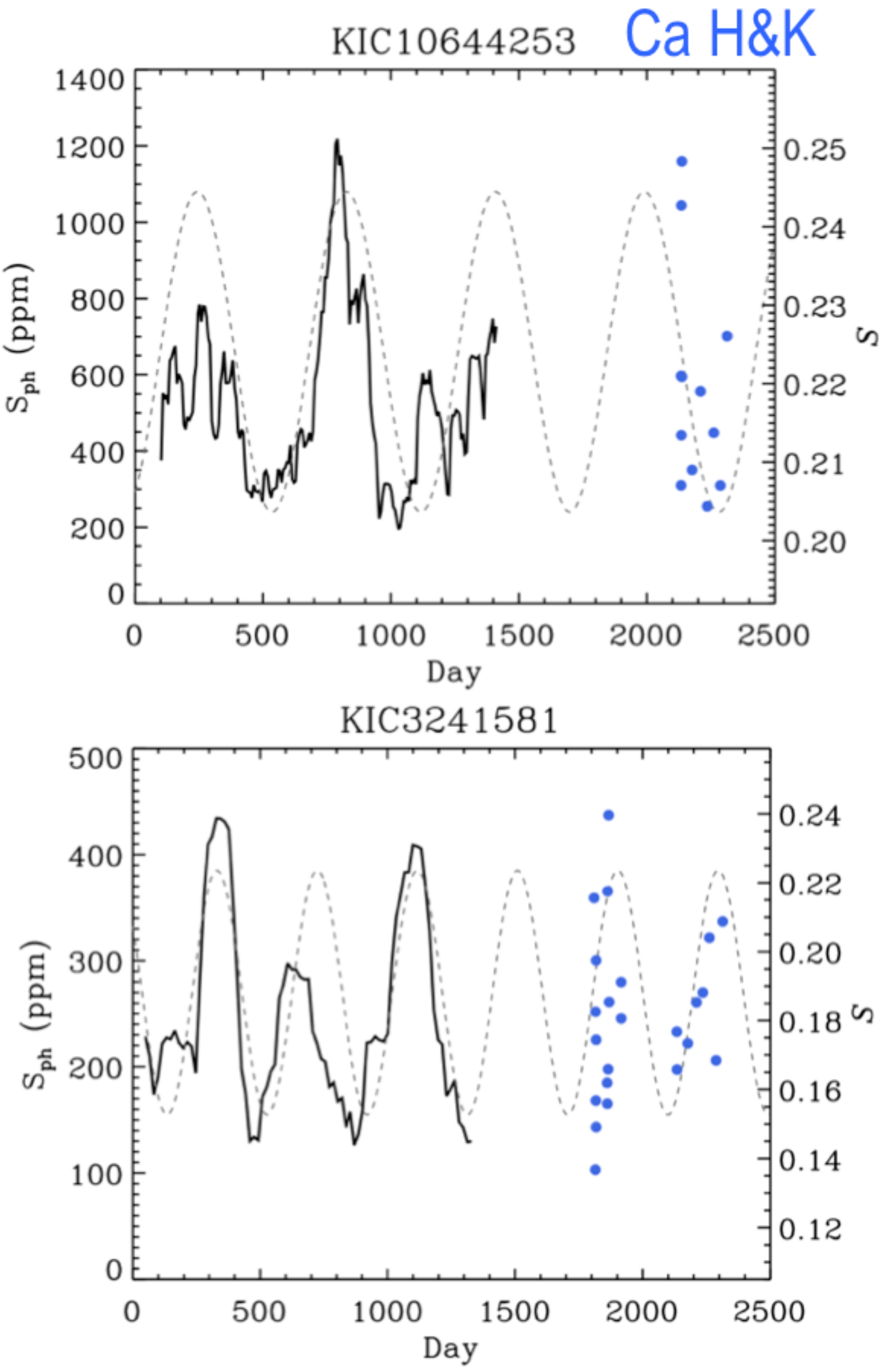}
\caption{Time evolution of the $S_{\rm{ph}}$ from \cite{2016A&A...589A.118S} obtained from the \emph{Kepler} photometry (black continuous line) and Chromospheric S index obtained form the spectroscopic follow up observations obtained from the Hermes/Mercator telescope (blue dots) for two solar analogues, KIC~10644253 and KIC~3241581. The dashed curve represent the potential magnetic modulation plotted to guide the eye between the two sets of observations.
}
\label{fig-6}       % Give a unique label
\end{figure}

Follow up observations have been engaged in the Hermes/Mercator telescope during the last two years for all these targets. An example is shown in figure~\ref{fig-6} for KIC~10644253 and KIC~3241581. The chromospheric S index seems to follow the magnetic modulation depicted form the \emph{Kepler} $S_{\rm{ph}}$ proxy. However, longer follow up observations will be required to confirm the periodicity of the magnetic activity cycle of those stars.
%++++++++++++++
%++++++++++++++

%-----------------------------------------------------------------------
\section{Conclusions}
The global helioseismology working group (4.1) in the SPACEINN project has been a success because of the quality and quantity of the work done as well as for the tools and datasets delivered to the community through the SPACEINN web portal. But the work of this team will not finish at the end of 2016 with the official end of the project. WP4.1 is still working on several scientific papers (e.g. Broomhall et al., Corsaro et al., Salabert et al.) and it will continue working together in some other on-going analysis of helioseismic data. It is also important to mention that the deliverables provided go beyond what was originally proposed in the project. WP4.1 has also been a success reinforcing the synergies between helio- and asteroseismology with a continuous transfer from one to the other, in particular, on the studies of solar analogue stars, which are allowing us to better understand the Sun compared to its siblings. 
\label{sec-con}

%-----------------------------------------------------------------------
% BibTeX or Biber users please use (the style is already called in
% the class, ensure that the "woc.bst" style is in your local directory)
% \bibliography{name or your bibliography database}

\bibliography{./GarciaRA_Biblio}

%\begin{thebibliography}{}
    % Format for Journal Reference Journal Author, Journal \textbf{Volume}, page numbers (year)
  %Book Author, \textit{Book title} (Publisher, place, year) page numbers
  % etc
%\end{thebibliography}

%***********************************************************************
\end{document}